\documentclass[a4paper,10pt]{article}
\usepackage[utf8x]{inputenc}

\title{A Superspace Formulation of The BV Action for Higher Derivative Theories}
\author{ Mir Faizal \\
Department of Mathematics,
Durham University\\
South Road,
Durham DH1 3LE\\
faizal.mir@durham.ac.uk\\ \\
Mozzam Khan\\
Department of Mathematics,
King's College London\\
Strand,
London WC2R 2L\\
mozzam.khan@kcl.ac.uk}
\begin{document}

\maketitle

\begin{abstract}
We first analyze the anti-BRST and double BRST structures of a certain higher derivative theory that has
been known to possess BRST symmetry associated with its higher derivative structure.
We discuss the invariance of this theory under shift symmetry in the Batalin Vilkovisky (BV) formalism.  We show that the action for this theory
can be written in a manifestly extended BRST invariant manner in  superspace formalism using one Grassmann coordinate.
 It can also be written in a manifestly extended BRST invariant manner and on-shell  manifestly extended anti-BRST  invariant manner in
superspace formalism using two Grassmann coordinates.
\end{abstract}



\section{Introduction}
Higher derivative theories have better renormalisation properties than the conventional
theories and thus have been thoroughly studied \cite{1a}-\cite{2a}.
 However, these higher derivative theories possess negative norm states or ghost states,
which break the unitarity of these theories and render them inconsistent as quantum field
theories \cite{3a}-\cite{4a}.  Many attempts using some kind of superselection rules or
 subsidiary conditions have been imposed to overcome this problem \cite{5a}-\cite{6a}.  But so far no one
has been able to give a general method to deal with these ghost states. However ghosts also occur in
conventional gauge theories  as Faddeev-Popov ghosts.  In conventional gauge theories the gauge-fixing term,
 the ghost term and the original classical Lagrangian density are invariant under symmetries called the BRST and the anti-BRST symmetries.
 This BRST or anti-BRST invariance of the total Lagrangian density is used to remove all the negative norm states and get a unitary theory
 \cite{7a}-\cite{8a}.

Recently BRST symmetry in higher derivative theories has been studied \cite{9a}.  This BRST symmetry is not due to gauge
fixing but is an intrinsic feature of these higher derivative theories. Thus in \cite{9a}, it was shown that even a simple higher derivative scalar
 field theory, along with suitably chosen ghost term, possesses BRST symmetry.  In this paper, we will  show that it also possesses
an anti-BRST symmetry.  In fact, we will show that it possess a double BRST symmetry.

We will then study the extended BRST and extended anti-BRST symmetries for this theory in the Batalin Vilkovisky (BV) formalism \cite{10a}-\cite{11a}.
 BV formalism has been studied, in the context of both the extended BRST, and the extended anti-BRST symmetries, in conventional gauge theories \cite{12a}-\cite{13a}.
 Furthermore, a superspace formalism for the BV formalism in conventional gauge theories is also well understood \cite{14a}-\cite{15a}. This  motivates the
study
of the higher derivative theories in the superspace BV formalism.

\section{BRST and Anti-BRST Invariant Lagrangian Density}
The higher derivative theory that we will analyze in this paper is a scalar field theory with suitable added ghost terms,
that is known to possess an intrinsic BRST symmetry.  We thus take the following $d$ dimensional action  
\begin{equation}
S=\int d^{d}x\mathcal{L}. \label{1a}
\end{equation}
 Here the Lagrangian density $\mathcal{L}$, is given by
\begin{equation}
\mathcal{L}=\frac{1}{2}\mathcal{O}\phi \mathcal{O}\phi + \overline{c}\mathcal{O} c, \label{2a}
\end{equation}
where $c$ is a ghost field, $\overline{c}$  is an anti-ghost field, $\phi$ is a real scalar field and 
\begin{equation}
\mathcal{O}=\prod_{i}^{N}[\nabla^{2}+m_{i}^{2}].\label{3a}
\end{equation}
This Lagrangian density  can then be rewritten using an auxiliary field $F$ as
\begin{equation}
\mathcal{L}=F\mathcal{O}\phi-\frac{1}{2}F^{2} + \overline{c}\mathcal{O} c. \label{4a}
\end{equation}
The  Lagrangian density given by Eq $(\ref{4a})$ is known to possess a BRST symmetry \cite{9a}, as it is  invariant under the following BRST transformations
\begin{eqnarray}
\delta \phi&=&c,\nonumber\\
\delta \overline{c}&=&-F,\nonumber\\
\delta c&=&0,\nonumber\\
\delta F&=&0.
\end{eqnarray}
We note that this  Lagrangian density given by Eq $(\ref{4a})$ is also invariant under the following anti-BRST transformations 
\begin{eqnarray}
\overline{\delta}\phi&=&\overline{c},\nonumber \\
\overline{\delta c}&=&0,\nonumber \\
\overline{\delta}c&=&F, \nonumber \\
\overline{\delta}F&=&0,\label{6a}
\end{eqnarray}
and so it can be written as
\begin{eqnarray}
\nonumber
\mathcal{L}&=&\overline{\delta}[c(\mathcal{O}\phi -\frac{1}{2}F)] \nonumber \\
&=&-\delta[\overline{c}(\mathcal{O}\phi-\frac{1}{2}F)] \nonumber \\
&=&\frac{1}{2}\overline{\delta}\delta[\phi \mathcal{O}\phi-c\overline{c}] \nonumber \\
&=&-\frac{1}{2}\delta\overline{\delta}[\phi \mathcal{O}\phi - c\overline{c}]. \label{7aab}
\end{eqnarray}
In this section we reviewed the BRST invariance of the Lagrangian density for higher derivative scalar field theory with suitable added ghost terms.
In the next section we will analyze the extended BRST invariance of this Lagrangian density.
\section{Extended BRST Lagrangian Density}
Extended Lagrangian density is obtained by requiring the original Lagrangian density to be invariant under both the original BRST transformations and the shift
transformations of the original fields.  Thus we  we make the following shift transformations
\begin{eqnarray}
\phi &\to& \phi - \tilde{\phi}, \nonumber \\
c &\to& c - \tilde{c}, \nonumber \\
\overline{c} &\to& \overline{c} - \tilde{\overline{c}},\nonumber\\
F &\to&  F - \tilde{ F}.
\end{eqnarray}
The extended BRST invariant formulation of the BV action is obtained by requiring the Lagrangian density
\begin{equation}
\tilde{\mathcal{ L}}=\mathcal{L}(\phi - \tilde{\phi}, c- \overline{\tilde c}, \overline{c}-\overline{\tilde c}, F- \tilde F),  \label{9a}
\end{equation}
to be invariant under the original BRST transformations and the shift transformations.
Thus the Lagrangian density given by Eq. $(\ref{9a})$ is invariant under the following  extended BRST symmetry with the transformations
\begin{eqnarray}
\delta\phi=\psi, && \delta \tilde\phi=(\psi- (c- \tilde c)),\nonumber\\
\delta c =\epsilon, && \delta \tilde{c}=\epsilon,\nonumber \\
\delta \overline c = \epsilon,&&  \delta \tilde{\overline c}=(\overline \epsilon + (F-\overline F)), \nonumber\\
\delta F =\rho, && \delta\overline F=\rho.
\end{eqnarray}
Here, $\psi, \epsilon, \overline \epsilon $ and $\rho$ are the ghost fields associated with the shift symmetries for $\phi, c, \overline {c}$ and $F$ respectively
with their corresponding BRST transformation vanishing:
\begin{eqnarray}
\delta\psi &=&0,\nonumber\\
\delta\epsilon &=&0,\nonumber \\
\delta \tilde{\epsilon}&=&0,\nonumber\\
\delta \rho&=&0.
\end{eqnarray}
We will further need to add anti-fields with opposite parity, corresponding to each of the fields with the following BRST transformations:
\begin{eqnarray}
\delta \phi^{*}&=&-b,\nonumber\\
\delta c^{*}&=&-B,\nonumber\\
\delta \overline c^{*}&=&-\overline{B}, \nonumber \\
\delta F^{*}&=&-\overline{b}.
\end{eqnarray}
The BRST transformations of the new auxiliary fields also vanishes:
\begin{eqnarray}
\delta b&=0,\nonumber \\
\delta B&=0,\nonumber \\
\delta \overline B&=0,\nonumber \\
\delta \overline{b}&=0.\label{3b}
\end{eqnarray}
We can now gauge fix the shift symmetry such that the tilde fields vanish and thus recover our original theory.
 This can be achieved by choosing the following Lagrangian density
\begin{eqnarray}
\tilde {\mathcal{L}}&=&-b\tilde{\phi}-\phi^{*}(\psi-(c- \tilde{c}))
-\overline{B} \tilde{c} + \overline{c}^{*}\epsilon \nonumber \\ &&
+B\tilde{\overline{c}}-c^{*}(\overline{\epsilon}+(F-\tilde{F}))+\overline{b}\tilde{F}+F^{*}\rho. 
\end{eqnarray}
Now  all the tilde fields will vanish upon integrating out the auxiliary fields $b, B, \overline{B}$ and $\overline{b}$.
 This Lagrangian is also invariant under the original BRST transformation and the shift transformations.
 Along with this Lagrangian density we have the original Lagrangian density, which is only a function of the original fields.
 So we can define  $\Psi=-[\overline{c}(\mathcal{O}\phi-F/2)]$, and then  by using Eq. $(\ref{7aab})$, we can write the original Lagrangian density as
\begin{equation}
\mathcal{L}=\delta\Psi.  \label{8a}
\end{equation}
This gives us
\begin{eqnarray}
\mathcal{L}&=&\delta\phi \frac{\delta \Psi}{\delta\phi}+\delta c \frac{\delta \Psi}{\delta c}+\delta \overline{c}
\frac{\delta \Psi}{\delta \overline{c}}+\delta F\frac{\delta \Psi}{\delta F}\nonumber \\ 
&=&-\frac{\delta \Psi}{\delta\phi}\psi+\frac{\delta \Psi}{\delta c}\epsilon +\frac{\delta \Psi}{\delta \overline{c}}
\overline{\epsilon}-\frac{\delta \Psi}{\delta F}\rho. \label{4b}
\end{eqnarray}
Again if we integrate out the fields setting the tilde to zero, we have
\begin{eqnarray}
\mathcal{L}_{\rm{tot}}&=&\tilde {\mathcal{L}}+\mathcal{L}\nonumber\\
&=&\phi^{*}c-c^{*}F-(\phi^* +\frac{\delta \Psi}{\delta F})\psi\nonumber \\
&&+(\overline{c}^{*} + \frac{\delta \Psi}{\delta c})\epsilon-(c^{*} - \frac{\delta \Psi}{\delta \overline{c}})\overline{\epsilon}+(F^{*}-
 \frac{\delta \Psi}{\delta F})\rho.
\end{eqnarray}
If we now integrate out the ghosts associated with the shift symmetry, it leads to the identification
\begin{eqnarray}
\phi^{*}&=&-\frac{\delta\Psi}{\delta\phi},\nonumber\\
\overline{c}^{*}&=&-\frac{\delta \Psi}{\delta c},\nonumber\\
c^{*}&=&\frac{\delta \Psi}{\delta \overline{c}},\nonumber\\
F^{*}&=&\frac{\delta \Psi}{\delta F}.
\end{eqnarray}
Thus we get
\begin{eqnarray}
\phi^{*}&=& \mathcal{O}\overline{c},\nonumber\\
\overline{c}^{*}&=&0,\nonumber\\
c^{*}&=&-\mathcal{O}\phi+ \frac{F}{2},\nonumber\\
F^{*}&=&\frac{\overline{c}}{2}. \label{7b}
\end{eqnarray}
With these identifications we obtain the BV action of the theory.
\section{Extended BRST  Superspace }
In this section we will convert the results of the previous section into a superspace formalism.
We now consider the anti-commutating parameter $\theta$ and define the following superfields
\begin{eqnarray}
\varphi(x,\theta)&=&\phi+\theta \psi,\nonumber\\
\tilde{\varphi}(x,\theta)&=&\tilde{\phi}+\theta(\psi-(c-\tilde{c})),\nonumber\\
\chi(x,\theta)&=&c+\theta \epsilon,\nonumber \\
\tilde{\chi}(x,\theta)&=&\tilde{c}+\theta\epsilon,\nonumber \\
\overline{\chi}(x,\theta)&=&\overline{c}+\theta\overline{\epsilon},\nonumber\\
\tilde{\overline{\chi}}(x,\theta)&=&\tilde{\overline{c}}+\theta(\overline{\epsilon}+(F-\tilde{F})), \nonumber\\
f(x,\theta)&=& F+\theta \rho,\nonumber\\
\tilde{f}(x,\theta)&=&\tilde{F}+\theta \rho.
\label{8ba}
\end{eqnarray}
We also define the following anti-superfields
\begin{eqnarray}
\tilde{\varphi}^{*}(x,\theta)&=&\phi^{*}-\theta b,\nonumber\\
\tilde{\chi}^{*}(x,\theta)&=&c^{*}-\theta B, \nonumber \\
\tilde{\overline{\chi}}^{*}(x,\theta)&=&\tilde{c}^{*}-\theta \overline{B},\nonumber \\
\tilde{f}^{*}(x,\theta)&=& F^{*}-\theta \overline{b}.
\end{eqnarray}
Thus, from these superfields and anti-superfields,  we get
\begin{eqnarray}
\frac{\partial}{\partial\theta}  \tilde{\varphi^{*}}\tilde{\varphi}&=&-b\tilde{\phi}-\phi^{*}(\psi-(c-\tilde{c})),\nonumber\\ 
\frac{\partial}{\partial \theta}  \tilde{\overline{\chi}^{*}}\tilde{\chi}&=&-\overline{B}\tilde{c}+\overline{c}^{*}\epsilon,\nonumber \\
-\frac{\partial}{\partial \theta}  \tilde{\overline{\chi}}\tilde{\chi}^{*}&=&B\tilde{\overline{c}}-c^{*}(\overline{\epsilon}+(F-\tilde{F})),\nonumber\\ 
-\frac{\partial}{\partial \theta}\tilde{f}^{*}\tilde{f}&=&\overline{b}\tilde{F}+F^{*}\rho.
\end{eqnarray}
Thus the Lagrangian density given by Eq. $(\ref{3b})$  can be written, in this superspace formalism, as
\begin{equation}
\tilde{\mathcal{L}}=\frac{\partial}{\partial {\theta}}(\tilde{\varphi}^{*}\tilde{\varphi}+
\tilde{\overline{\chi}}^{*}\tilde{\chi}-\tilde{\overline{\chi}}\tilde{\chi}^{*}-\tilde{f}^{*}\tilde{f}). \label{1c}
\end{equation}
Being the $\theta$ component of a superfield this is manifestly invariant under the extended BRST transformation.  The gauge fixing Lagrangian 
density  for the original
 symmetry can also be written in this formalism by defining $\Phi$ as
\begin{equation}
\Phi = \Psi + \theta \delta \Psi.
\end{equation}
Thus we have
\begin{equation}
\Phi=\Psi+\theta(-\frac{\delta \Psi}{\delta\phi}\psi+\frac{\delta \Psi}{\delta c}\epsilon
+ \frac{\delta \Psi}{\delta \overline{c}}\overline{\epsilon}-\frac{\delta\Psi}{\delta F}\rho). \label{2c}
\end{equation}
So we can now write the original gauge-fixing Lagrangian density in the superspace formalism as
\begin{equation}
\mathcal{L}=\frac{\partial \Phi}{\partial \theta}. \label{4c}
\end{equation}
Once again, being the $\theta$ component of a superfield, this is manifestly invariant under the extended BRST transformation.  The complete Lagrangian density
 can now be written as
\begin{eqnarray}
\tilde{\mathcal{L}}_{\rm{tot}}&=&\tilde{\mathcal{L}}+\mathcal{L}\nonumber \\
&=&\frac{\partial}{\partial\theta} (\tilde{\varphi}^{*}\tilde{\varphi}+\tilde{\overline{\chi}}^{*}\tilde{\chi}-\tilde{\overline{\chi}}\tilde{\chi}^{*}-
\tilde{f}^{*}\tilde{f})+\frac{\partial{\Phi}}{\partial \theta}.
\end{eqnarray}
This Lagrangian density is manifestly invariant under the BRST symmetry, after elimination of the auxiliary and ghost fields associated with the shift symmetry.
\section{Extended Anti-BRST Lagrangian Density}
In the previous sections we analysed the extended BRST symmetry for the Lagrangian density of higher derivative scalar field theory with suitably chosen
ghost terms.  Now we will discuss the extended anti-BRST symmetry of this theory.  The original and shifted fields obey the extended anti-BRST transformations,
\begin{eqnarray}
\overline{\delta}\tilde{\phi}=\phi^{*}, && \overline{\delta}\phi= \phi^{*}+(c-\tilde{\overline{c}}), \nonumber\\
\overline{\delta}\tilde{c}=\overline{c}^{*}, && \overline{\delta}c= c^{*}+(F-\overline{F}), \nonumber\\
\overline{\delta}\tilde{\overline{c}}=\overline{c}^{*}, && \overline{\delta}\overline{c}=\overline{c}^{*},\nonumber\\
\overline{\delta}\tilde{F}=F^{*}, && \overline{\delta}F= F^{*}.
\end{eqnarray}
The ghost fields associated with the shift symmetry have the following extended anti-BRST transformations,
\begin{eqnarray}
\overline{\delta}\psi&=&b+(F-\tilde{F}),\nonumber\\
\overline{\delta} \epsilon&=&B,\nonumber\\
\overline{\delta}\overline{\epsilon}&=&\overline{B},\nonumber\\
\overline{\delta} \rho&=&\overline{b}.
\end{eqnarray}
and the extended anti-BRST transformations of the anti-fields of the auxiliary fields associated with the shift symmetry vanishes,
\begin{eqnarray}
\overline{\delta }b =0, && \overline{\delta} \phi^* =0,\nonumber\\
\overline{\delta} B=0, &&\overline{\delta} c^{*}=0,\nonumber\\
\overline{\delta} \overline{B}=0, &&\overline{\delta} \overline c^{*}=0,\nonumber\\
\overline{\delta }\overline{b}=0, && \overline{\delta} F^{*}=0.
\end{eqnarray}
For the Lagrangian density which is both BRST and anti-BRST invariant, it follows that it  must also be invariant under the extended anti-BRST transformation
 at least on-shell where the transformations reduce to anti-BRST transformations.
\section{Extended Anti-BRST Superspace }
To write a Lagrangian density that is manifestly invariant both under extended BRST transformations and under extended anti-BRST transformations
 on-shell in superspace formalism, we will have to define superfields in an analogous way to what was done in the previous sections, but with two anti-commutating
parameters, namely $\theta$ and $\overline{\theta}$.  So we can define the following superfields,
\begin{eqnarray}
\varphi(x,\theta,\overline{\theta})&=&\phi+\theta \psi+\overline{\theta}(\phi^{*}+(\overline{c}-\tilde{\overline{c}}))+\theta \overline{\theta}(b+ (F-\tilde{F})),
\nonumber\\
\tilde{\varphi}(x,\theta,\overline{\theta})&=&\tilde{\phi}+\theta(\psi-(c-\tilde{c}))+\overline{\theta}\phi^{*}+\theta \overline{\theta}b,
\nonumber\\
\chi(x,\theta,\overline{\theta})&=&c+\theta \epsilon
+\overline{\theta}(c^{*}+(F-\tilde{F}))+\theta\overline{\theta}B,
 \nonumber\\
\tilde{\chi}(x,\theta,\overline{\theta})&=&\tilde{c}+\theta \epsilon+\overline{\theta}c^{*}+\theta\overline{\theta}B,
\nonumber\\
\overline{\chi}(x,\theta,\overline{\theta})&=&\overline{c}+\theta\overline{\epsilon}+\overline{\theta} \overline{c}^* +\theta \overline{\theta}\overline{B},
\nonumber\\
\tilde{\overline{\chi}}(x,\theta,\overline{\theta})&=&\tilde{\overline{c}}+\theta(\overline{\epsilon}+(F-\tilde{F})) + \overline{\theta}\overline{c}^{*}
+\theta \overline{\theta}\overline{B}. \label{7c}
\end{eqnarray}
Now we have
\begin{eqnarray}
 -\frac{1}{2}\frac{\partial}{\partial \overline{\theta}}\frac{\partial}{\partial \theta} \tilde{\varphi}\tilde{\varphi}&=&-b\tilde{\phi}
-\phi^{*}(\psi-(c-\tilde{c})),\\
\frac{\partial}{\partial \overline{\theta}}\frac{\partial}{\partial \theta}\tilde{\chi}\tilde{\overline{\chi}}&=& -\overline{B}\tilde{c}
+\overline{c}^{*}\epsilon +B\tilde{\overline{c}}-c^{*}(\overline{\epsilon}+(F-\tilde{F})).
\end{eqnarray}
Consequently we can write
\begin{eqnarray}
 \tilde{\mathcal{L}}&=&\frac{\partial}{\partial \overline{\theta}}\frac{\partial}{\partial \theta}
 (-\frac{1}{2}\tilde{\varphi}\tilde{\varphi}+\tilde{\chi}\tilde{\overline{\chi}})\nonumber \nonumber \\ &=&
-b\tilde{\phi}-\phi^{*}(\psi-(c-\tilde{c}))-\overline{B}\tilde{c}+\overline{c}^{*}\epsilon\\ 
&&+B\tilde{\overline{c}}-c^{*}(\overline{\epsilon}+(F-\tilde{F})).
\end{eqnarray}
Being the $\theta \overline{\theta}$ component of a superfield, this gauge-fixing Lagrangian density is manifestly invariant under extended BRST and anti-BRST
 transformations. Furthermore, we now define
\begin{equation}
\Phi(x,\theta,\overline{\theta})=\Psi+\theta \delta \Psi+ \overline{\theta}\overline{\delta}\Psi
+\theta \overline{\theta} \delta \overline{\delta}\Psi. \label{3d}
\end{equation}
The component of $\theta \overline{\theta}$ can be made to vanish on-shell and therefore the Lagrangian density for the original
 fields can be written as
\begin{equation}
\mathcal{L}=\frac{\partial}{\partial \theta}(\delta(\overline{\theta}) \Phi(x, \theta, \overline{\theta})). \label{7d}
\end{equation}
This Lagrangian density is not only manifestly invariant under extended BRST transformations but also invariant under extended anti-BRST transformations
 on-shell.  The complete Lagrangian density can therefore be written as
\begin{eqnarray}
\mathcal{L}_{\rm{tot}}&=&\tilde{\mathcal{L}} +\mathcal{L}\nonumber\\
&=&\frac{\partial}{\partial \overline{\theta}} \frac{\partial}{\partial \theta}(-\frac{1}{2}\tilde{\varphi}\tilde{\varphi}+
 \tilde{\chi}\tilde{\overline{\chi}}) + \frac{\partial}{\partial \theta}(\delta(\overline{\theta})\Phi(x,\theta,\overline{\theta}))\nonumber\\
&=&-b\tilde{\phi}-\overline{B}\tilde{c}+B\tilde{\overline{c}} - (\phi^{*}+\frac{\delta \Psi}{\delta \phi})\psi
+\phi^{*}(c-\tilde{c})-c^{*}(F-\tilde{F}) \nonumber\\
&& +(\overline{c}^{*}
 + \frac{\delta \Psi}{\delta c})\epsilon-(c^{*}-
\frac{\delta \Psi}{\delta \overline{c}})\overline{\epsilon}. \label{8d}
\end{eqnarray}
By integrating out the auxiliary fields the tilde fields are made to vanish and
 by integrating out the ghost fields for the shift symmetry we will get explicit expressions for the
 antifields.
It may be noted as $F$ and $\tilde{F}$ are auxiliary fields so we can redefine them as $F-\tilde{F}\to F$.
Then the combination $(F+\tilde{F})$ can then be integrated out and absorbed into the normalization constant.
  Thus we have obtained a  Lagrangian density in superspace formalism which
is manifestly BRST invariant and also manifestly anti-BRST invariant on-shell.
\section{Conclusion}
We have analysed higher derivative scalar field theory with suitably added ghost terms in the BV formalism.
 We have shown that we can write the Lagrangian density for a higher derivative scalar field theory with
 suitably added ghost terms in a manifestly BRST invariant and manifestly on-shell anti-BRST  invariant way using the superspace formalism.
We can apply this formalism to higher derivative Lagrangian density for Yang-Mills theories and higher derivative Lagrangian density for gravity.
In \cite{9a}, intrinsic BRST symmetry for both these theories has already been analysed. It would now be natural to extend this present work to these theories.

\end{document}